\documentclass{article}

\usepackage{epsf}
\usepackage{cite}
\usepackage{latexsym}
\usepackage{amssymb}
\usepackage{amsmath}
\begin{document}

\title{Renormalized stress tensor in one-bubble spacetimes}
\author{Xavier Montes\\{\sl IFAE, Edifici C,
  Universitat Aut{\`o}noma de Barcelona},\\ 
{\sl E-08193 Bellaterra, Spain}\\ \\
\hskip 8.5cm {\small UAB-FT-461}}
\date{}
\maketitle
\begin{abstract}
  We compute the two-point function and the renormalized expectation
  value of the stress tensor of a quantum field interacting with a
  nucleating bubble. Two simple models are considered. One is the
  massless field in the Vilenkin-Ipser-Sikivie spacetime describing
  the gravitational field of a reflection symmetric domain wall. The
  other is vacuum decay in flat spacetime where the quantum field only
  interacts with the tunneling field on the bubble wall. In both cases
  the stress tensor is of the perfect fluid form.  The assymptotic
  form of the equation of state are given for each model.  In the VIS
  case, we find that $p=-(1/3)\rho$, where the energy density $\rho$
  is dominated by the gradients of supercurvature modes.
\end{abstract}

\section{Introduction}
\begin{figure}[t]
\centering

\leavevmode\epsfysize=5cm \epsfbox{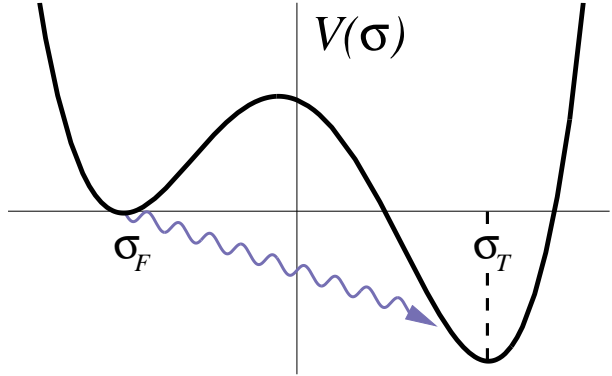}\\[3mm]
\begin{quote}
{\sl 
\caption[fig1]{\label{potencial} Assumed shape for the potential of the
  tunneling field. It has a local minimum which corresponds to the false
  vacuum at $\sigma_F$ and a global minimum, the true vacuum, at $\sigma_T$.
The bounce corresponds to the Euclidean evolution of the tunneling field under
the barrier.}}
\end{quote}
\end{figure}

The problem of the quantum state of a nucleating bubble has been addressed in
the literature several times\cite{coleman,rubakov,VV,MT,TMdecay}. The results
relevant for our discussion can be summarized as follows. We have a
self-interacting scalar field $\sigma$ (the tunneling field) described by the
lagrangian
\begin{equation}
{\mathcal L}_\sigma = -\frac{1}{2}\partial_\mu\sigma\partial^\mu\sigma-V(\sigma)
\end{equation}
where $V(\sigma)$ has a local (metastable) minimum at some value
$\sigma_F$ and a global one at $\sigma_T$ (see Fig.~\ref{potencial}).
The bubble nucleation can be pictured as the evolution of the $\sigma$
field in imaginary time. The solution of the corresponding Euclidean
time equation which interpolates between the false vacuum at spacetime
infinity and the true vacuum inside the bubble is called the bounce.
In the absence of gravity, vacuum decay is dominated by the $O(4)$
symmetric bounce solution \cite{colemanmin}. So we shall write the
tunneling field as a function of $\tau\equiv(T_E^2+{\bf X})^{1/2}$
alone,
\begin{equation}
\sigma = \sigma_0(\tau),
\end{equation}
where $(T_E,{\bf X})$ are Cartesian coordinates in Euclidean space. The
solution describing the bubble after nucleation is given by the analytic
continuation of the bounce to Minkowski time $T$ through the substitution
$T_E=-iT$. Then, the bubble solution depends only on the Lorentz
invariant quantity $({\bf X}^2-T^2)^{1/2}$, where $(T,{\bf X})$ are the usual
Minkowski coordinates. 

If there are quantum fields interacting with the tunneling field, their state
will be significantly affected by the change of vacuum state.  Pioneering
investigations of this matter were carried out by Rubakov \cite{rubakov} and
Vachaspati and Vilenkin \cite{VV}. These latter authors considered a model of
two interacting scalar fields $\sigma$ and $\Phi$, and found the quantum state
for $\widehat{\bf \Phi}$ (the quantum counterpart of $\Phi$) by solving its
functional Scr{\"o}dinger equation. In order to find a solution, they impose
as boundary conditions for the wave function $\Psi(\tau;\Phi]$ regularity
under the barrier and the tunneling boundary condition (see \cite{VV} for
details). They found that the quantum state must be SO(3,1) invariant.

A somewhat different approach was pursued later by Sasaki and Tanaka
\cite{MT}.  They carried out a refinement of the method for
constructing the WKB wave function for multidimensional systems, first
introduced by Banks, Bender and Wu \cite{banks} and extended to field
theory by Vega, Gervais and Sakita \cite{vega}, and obtained the so
called quasi-ground state wave function. The quasi-ground state wave
function is a solution of the time independent functional
Schr{\"o}dinger equation to the second order in the WKB approximation
which is sufficiently localized at the false vacuum so that it would
be the ground state wave functional it there were no tunneling. They
also found that the state must be SO(3,1) invariant.

Moreover, general arguments, due to Coleman \cite{gencol}, suggest
that the decay must be SO(3,1) invariant. If not, the infinite volume
Lorentz group will make the nucleation probability diverge. From a
practical point of view, therefore, it would be interesting to know to
what extent symmetry considerations alone can be used to determine the
quantum state after nucleations. As a first approach to this question
it will be useful to compute the two-point function and the
renormalized expectation value of the stress tensor in a SO(3,1)
invariant quantum state for two simples models of one-bubble
spacetimes.
  
\section{General Formalism}

Our aim is to study the quantum state of a field $\Phi$ described by
a Lagrangian of the general form
\begin{equation}
{\mathcal L}_\Phi=-\frac{1}{2}\partial_\mu\Phi\partial^\mu\Phi - 
\frac{1}{2}m(\sigma)^2 \Phi^2,\label{wish}
\end{equation}
where the mass term is due to the interaction of the field $\Phi$ with
a nucleating bubble. Working from the very beginning in the Heisenberg
picture, we will construct an SO(3,1) invariant quantum state for the
field $\widehat{\bf \Phi}$. After we will find its Hadamard two-point
function $G^{(1)}(x,x')\equiv \langle
0|\{\widehat{\bf\Phi},(x)\widehat{\bf\Phi}(x')\}|0\rangle$, and we
will check whether it is of the Hadamard form
\cite{hadamardA,hadamardB,hadamardC,waldkay}.  Loosely speaking, a
Hadamard state can be described\footnote{For a more precise definition
  of Hadamard states see \cite{waldkay}.} as a state for which the
singular part of $G^{(1)}(x,x')$ takes the form
\begin{equation}
  G^{(1)}_{\rm sing}(x,x') = \frac{u}{\sigma}+v\log(\sigma),
\end{equation}
where $\sigma$ denotes half of the square of the geodesic distance between $x$
and $x'$, and $u$ and $v$ are smooth functions that can be expanded as a power
series in $\sigma$, at least for $x'$ in a small neighborhood of $x$. Hadamard
states are considered physically acceptable because for them the
point-splitting prescription gives a satisfactory definition of the
expectation value of the stress-energy tensor. After clarifying the singular
structure of $G^{(1)}(x,x')$, we will use the point-splitting formalism
\cite{christensen,waldreg,folacci,dorca} to compute the renormalized expectation
value of the energy-momentum tensor in this quantum state.
Finally we will briefly discuss the applicability of a uniqueness theorem for
quantum states due to Kay and Wald \cite{waldkay}. 

\section{SO(3,1) coordinates}
In the present paper we will restrict ourselves to piecewise flat
spacetime. It proves very useful to use coordinates adapted to the
symmetry of the problem.  So we will coordinatize flat Minkowski space
using hyperbolic slices, which will embody the symmetry under Lorentz
transformations. We define the new coordinates $(t,r)$ (Milne
coordinates) by the equations
\begin{equation}
t \equiv (T^2-{\bf X}^2)^{1/2} \hskip 1cm r\equiv 
\mbox{tanh}^{-1} (|{\bf X}|/T),
\end{equation}
where $(T,{\bf X})$ are the usual Minkowski coordinates. In terms of these
coordinates, we have
\begin{equation}
ds^2=-dt^2+t^2d\Omega_{H_3},
\end{equation}
where 
\begin{equation}
  d\Omega_{H_3}=dr^2+\sinh^2r\,d\Omega_{S_2}
\end{equation}
is the metric on the unit 3-dimensional spacelike hyperboloid, and
$d\Omega_{S_2}$ is the line element on a unit sphere.

The above coordinates cover only the interior of the lightcone from the
origin. In order to cover the exterior, we will use the Rindler coordinates
\begin{equation}
\xi_R\equiv({\bf X}-T^2)^{1/2} \hskip 1cm \chi_R \equiv
\mbox{tanh}^{-1}(T/|{\bf X}|).
\end{equation}
In terms of this coordinates, the line element reads
\begin{equation}
ds^2 = d\xi_R^2 + g_{AB} dx^A dx^B = d\xi_R^2 + \xi_R^2\,d\Omega_{dS_3},
\end{equation}
where $g_{AB}$ is the metric on the $\xi_R=\mbox{ct.}$ hypersurfaces,
and $dS_3$ is the line element on a unit ``radius'' (2+1)-dimensional
de Sitter space,
\begin{equation}
d\Omega_{dS_3} = -d\chi_R^2 + \cosh^2\chi_R \,d\Omega_{S_2}.
\end{equation}   

The Milne and the Rindler coordinates are related by analytic
continuation,
\begin{equation}
\chi_R = r-i\pi/2 \hskip 1cm \xi_R =  i t. 
\end{equation}
Notice that $t$ is timelike inside the lightcone and becomes
spacelike after analytical continuation to the outside, whereas $r$ is
spacelike inside the lightcone but its analytical continuation is
time-like.
\begin{figure}[t]
\centering
\leavevmode\epsfysize=9cm \epsfbox{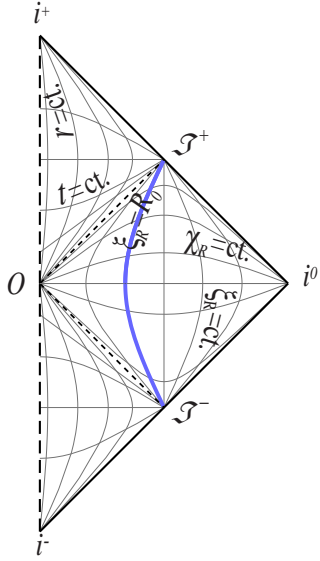}\\[3mm]

\begin{quote}
{\sl
\caption[fig2]{\label{minkowski} Conformal diagram of Minkowski spacetime. The
  Milne coordinates $(t,r)$ cover the region inside the lightcone
  emanating from the origin $O$. The Rindler coordinates
  $(\xi_R,\chi_R)$ cover the outside of this lightcone. The thicker
  solid line in the central diamond shaped region corresponds to the
  position of the bubble wall.}}
\end{quote}
\end{figure}

\section{Quantum state}

\begin{figure}[t]
\centering
\leavevmode\epsfysize=8cm \epsfbox{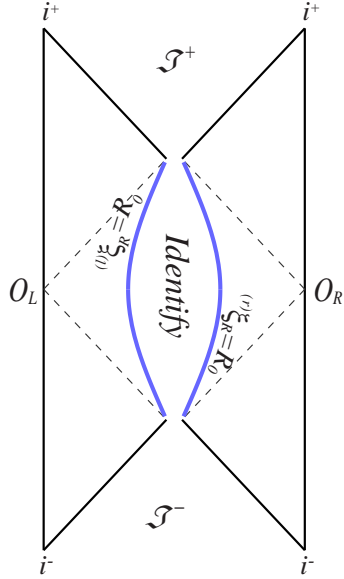}\\[3mm]
\begin{quote}
{\sl
\caption[fig3]{\label{VIS} Conformal diagram of the VIS spacetime. This
  spacetime, which corresponds to the global gravitational field of a
  reflection symmetric domain wall, is constructed by identifying two
  Minkowski spacetimes at some $\xi_R=R_0$.}}
\end{quote}
\end{figure}

Here we will consider two simple models. First we shall consider a
massless field living in the Vilenkin-Ipser-Sikivie spacetime\cite{VIS}. The
VIS spacetime represents the global gravitational field of a reflection
symmetric domain wall, and can be constructed by gluing two Minkowski spaces
at some $\xi_R=R_0$, the locus corresponding to the evolution of the bubble
wall (see Fig.~\ref{VIS}).  The second model we will study is a field which
interacts with the tunneling field only on the bubble wall.  For the tunneling
field, we will assume the thin bubble wall approximation. More general models
of the form (\ref{wish}) will be considered elsewhere\cite{next}.

The quantization will be performed in the ``Rindler wedges'' of these spaces,
because the hypersurfaces $\chi_R=\text{ct.}$ are Cauchy surfaces for the whole
spacetime.
  
\subsection{VIS model}

Here we consider a massless field living in a spacetime constructed by gluing
two Minkowski spaces at some $\xi_R=R_0$. We take Rindler coordinates in the
region outside the origin of the two pieces, using a Rindler patch for each
one. On each side, the Rindler coordinate $^{(l/r)}\xi_R$ , where the index
$l$ or $r$ refers to the left or right pieces, ranges from $^{(l/r)}\xi_R=0$
on the lightcone to some value $^{(l/r)}\xi_R=R_0$, where the two Minkowski
pieces are identified.  Defining $^{(l)}\xi_R = R_0e^{\eta}$ and $^{(r)}\xi_R
= R_0 e^{-\eta}$, we can coordinatize both pieces letting $\eta $ range from
$-\infty$ to $\infty$.  Then the line element outside the lightcone becomes
\begin{equation}
ds^2 = a(\eta)^2\left(d\eta^2-d\chi_R^2+\cosh^2\chi_Rd\,\Omega_2\right), 
\end{equation}   
where $a(\eta)=R_0e^{\eta}\theta(-\eta) + R_0e^{-\eta}\theta(\eta)$. Here
$\theta(x)$ is the Heaviside step function.  

In order to construct a quantum state, we expand the field operator
$\widehat{\bf\Phi}$ in terms of a sum over a complete set of mode
functions times the corresponding creation and annihilation operators,
\begin{equation}
  \widehat{\bf\Phi}=\sum_{plm}a_{plm}\Phi_{plm} + \mbox{h.c.},
\end{equation}
The mode functions $\Phi_{plm}$ satisfy the field equation
\begin{equation}
\Box\Phi_{plm}=0\label{eqmassless},
\end{equation}
where $\Box$ stands here for the four dimensional d'Alambertian operator in
the VIS spacetime. Taking the ansatz
\begin{equation}
\Phi_{plm} = \frac{F_p(\eta)}{a(\eta)}{\mathcal
    Y}_{plm}(\tilde x)
\end{equation}
where $\tilde x = (\chi_R,\Omega) $, $\Omega=(\theta,\varphi)$, equation
(\ref{eqmassless}) decouples into
\begin{align}
^{dS}\Box {\mathcal Y}_{plm} &= (p^2+1){\mathcal Y}_{plm}\label{eqY}\\
\left[-\frac{d^2}{d\eta^2}-2\delta(\eta)\right]F_p &= p^2 F_p.\label{eqF}
\end{align}
Here $^{dS}\Box$ stands for the covariant d'Alembertian on a (2+1) de
Sitter space. Equations (\ref{eqY})-(\ref{eqF}) have the
interpretation that ${\mathcal Y}_{plm}$ are massive fields living in a
(2+1) de Sitter space, with the mass spectrum given by the eigenvalues
of the Schr{\"o}dinger equation for $F_p$. Solving (\ref{eqF}), we
find that the spectrum has a continuous two-fold degenerate part for
$p^2>0$ and a bound state with $p^2=-1$ (a zero mode). If we let $p$
take positive and negative values, the normalized mode functions $F_p$
for $p^2>0$, which are the usual scattering waves, can be written as
\begin{equation}
F_p = \frac{1}{\sqrt{2\pi}}\left((e^{ip\eta}+\rho(p)
  e^{-ip\eta})\theta(-\mbox{sgn}(p)\,\eta)\ + 
\sigma(p)e^{ip\eta}\theta\left(\mbox{sgn}(p)\,\eta\right)\right),\label{scatt}
\end{equation}
where
\begin{align}
\rho(p) &= -\frac{1}{i|p|+1},\\
\sigma(p) &= \frac{i|p|}{i|p|+1}.
\end{align}
The normalized supercurvature mode $p^2=-1$ is given by
\begin{equation}
F_{,-1} = \frac{a(\eta)}{R_0}, 
\end{equation}
where the coma indicates that $-1$ refers to $p^2$ instead of $p$.
As we are interested in a SO(3,1) invariant state, the natural choice
for ${\mathcal Y}_{plm}$ are the positive frequency (2+1) Bunch-Davies
modes \cite{bunchdavies},
\begin{equation}
{\mathcal Y}_{plm}(\tilde x) = \sqrt{\frac{\Gamma(l+1+ip)\Gamma(l+1-ip)}{2}}
\frac{P_{ip-1/2}^{-l-1/2}(i \sinh \chi_R)}
{\sqrt{i \cosh \chi_R}}Y_{lm}(\Omega),
\end{equation}
where $Y_{lm}(\Omega)$ are the usual spherical harmonics. With this choice, it
is straightforward to show that the quantum state for $\widehat\Phi$ is SO(3,1)
invariant.
 
Now we proceed to compute the two-point Wightman function $G^{(+)}(x,x')$,
\begin{align}
G^{(+)}(x,x') &\equiv \langle0|\widehat{\bf\Phi}(x)\widehat{\bf\Phi}(x')
|0\rangle =\sum_{lm}\Phi_{-1,lm}(x)\overline{\Phi_{-1,lm}(x')}
\nonumber\\&+
\int_{-\infty}^{\infty}dp\sum_{lm}\Phi_{plm}(x)
\overline{\Phi_{plm}(x')}.
\end{align}
From now on we will suppose that the two points $x$ and $x'$ belong to the
Rindler wedge of the ``left Minkowski'' space, so we will omit the $(l)$ index
for notational simplicity. Direct substitution of the mode functions gives
\begin{align}
G^{(+)}(x,x') &= \frac{1}{R_0^2}\sum_{lm}{\mathcal Y}_{-1,lm}(\tilde
x)\overline{{\mathcal Y}_{-1,lm}(\tilde x')}\nonumber\\&+
\frac{1}{2\pi R_0^2}\int^{\infty}_{-\infty}dp
\left(\xi_M^{ip-1}\xi'_{M}{}^{-ip-1}\nonumber\right.\\&+\left.
\frac{1}{ip-1}\xi_M^{ip-1}\xi'_{M}{}^{ip-1}\right)\sum_{lm}
{\mathcal Y}_{plm}(\tilde x)\overline{{\mathcal
    Y}_{plm}(\tilde x')},
\end{align}
where we have defined $\xi_M=e^\eta=\xi_R/R_0$. The two-point function
$G^{(+)}$ is SO(3,1) invariant because is a sum of SO(3,1) invariant
terms.  Due to our choice of positive frequency modes, the $lm$ sums
correspond to the two-point Wightman functions in the Euclidean vacuum
for massive and a massless scalar fields living in $(2+1)$ de Sitter
spacetime. The (3+1)-dimensional Lorentz group SO(3,1) is the same as
the group of (2+1)-dimensional de Sitter transformations, so the
two-point functions are Lorentz invariant by
construction\footnote{\label{zeroinv}We will
  follow\cite{zeromodeA,zeromodeB,zeromodeC,jaumekirsten} to construct
  an SO(3,1) invariant state for the supercurvature massless mode with
  $p^2=-1$.}  (its explicit form is given below). $G^{(+)}$ also
depens on the quantity $\xi_R$.  This is a function of the interval in
Minkowski space time, so it is Lorentz invariant too.
\begin{figure}[t]
\centering
\leavevmode\epsfysize=6cm \epsfbox{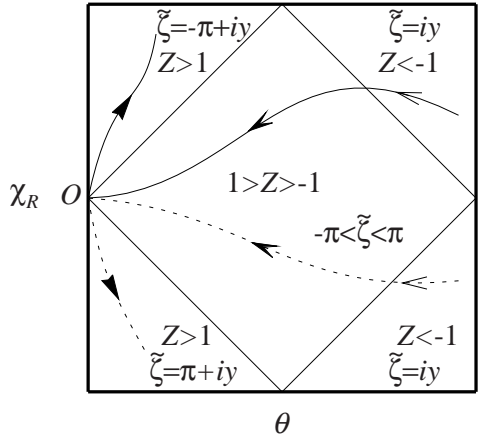}\\[3mm]
\begin{quote}
{\sl 
\caption[fig6]{\label{deSitter} Conformal diagram of a (2+1) de Sitter
  hyperboloid $\xi_R$=ct. Without loss of generality, we can
  take the point $\tilde x$ to lie in the ``origin'' O. Then $Z>1$ if $\tilde
  x'$ is timelike related with the origin, and $-1<Z<1$ if $\tilde x'$ is
  spacelike related with the origin and can be joined with it by means of a
  geodesic. If $Z<-1$, $\tilde x'$ is spacelike related with the origin but
  there are no geodesics connecting it with the origin. The function
  $\epsilon(x,x')$ is introduced in order to take into account the time
  ordering of those points which are timelike related.  If this is the case,
  it evaluates to $\varepsilon$ if $\chi_R>\chi_R'$ and to $-\varepsilon$ if
  $\chi_R<\chi_R'$, where $\varepsilon$ is a small positive value.  Two
  possible paths for $\chi'_R$ which pass through the origin are drawn (see
  discussion in Fig.~\ref{Zplane})}}
\end{quote}
\end{figure}

First we will compute the contribution of the continuum.  The $lm$ sum has
been explicitly computed \cite{sumaY},
\begin{align}
G^{(+)}_p(\tilde x,\tilde x') &\equiv \sum_{lm}{\mathcal Y}_{plm}\overline{{\mathcal
    Y}_{plm}} \nonumber\\&= \frac{\Gamma(1+ip)\Gamma(1-ip)}{(4\pi)^{3/2}
\Gamma(3/2)}
{}_2F_1\left[1+ip,1-ip;\frac{3}{2};{\frac{1+Z-i\epsilon}{2}}\right]
\label{sumcont},
\end{align}
where $Z(\tilde x,\tilde x')\equiv X^\mu(\tilde x)X_\mu(\tilde x') =
-\sinh\chi_R\sinh\chi_R'+
\cosh\chi_R\cosh\chi_R'\cos\widehat{\Omega\Omega'}$, which is
explicitly Lorentz invariant. Here $X^\mu(\tilde x)$ is the position
of the point $\tilde x$ in the (3+1) Minkowski space where the $(2+1)$
de Sitter space is embedded as a timelike hyperboloid. The function
$\epsilon(\tilde x,\tilde x')$ has been introduced to indicate at
which side of the cut the hypergeometric function should be
computed\footnote{The hypergeometric function in (\ref{sumcont}) has a
  branch cut along the real axis in the complex $Z$ plane from $Z=1$
  to $Z=\infty$.}.  It evaluates to $\varepsilon$ if $\tilde x$ and
$\tilde x'$ are timelike related and $\chi_R>\chi_R'$, to
$-\varepsilon$ if $\tilde x$ and $\tilde x'$ are timelike related and
$\chi_R<\chi_R'$, and vanishes if $\tilde x$ and $\tilde x'$ are
spacelike related, where $\varepsilon$ is a small positive constant
(see Fig.~\ref{deSitter}).  At the end of the calculation, we will
take the limit $\varepsilon\to 0$.  Introducing
$\cos\tilde\zeta\equiv-\tilde Z \equiv -Z+i\epsilon$, the two-point
function $G^{(+)}_p(\tilde x,\tilde x')$ can be compactly written as
\begin{equation}
G^{(+)}_p(\tilde x,\tilde x') =  \frac{1}{4\pi\sin\tilde\zeta}
\frac{\sinh p\tilde\zeta}{\sinh\pi p}.
\end{equation}
After performing the $p$ integration we obtain
\begin{align}
G^{(+)}_{\rm cont}(x,x') &= \frac{1}{8\pi^2\sigma}+\frac{1}{8\pi R_0^2}\left(
2\cot\tilde\zeta\left(\tilde\zeta - \mbox{arc
  tan}\frac{\sin\tilde\zeta}{\cos\tilde\zeta+\xi_M\xi_M'}\right)\right.
\nonumber\\
&+\left. \log\frac{(\xi_M\xi'_M)^2}
{\sin\tilde\zeta^2+(\cos\tilde\zeta+\xi_M\xi'_M)^2}\right).
\end{align}

\begin{figure}[t]
\centering
\leavevmode\epsfysize=4cm \epsfbox{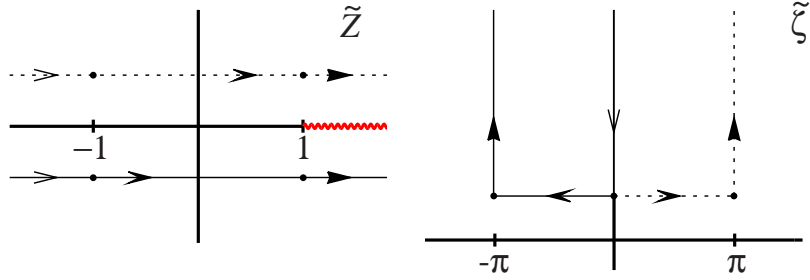}\\[3mm]
\begin{quote}
{\sl 
\caption[fig6]{\label{Zplane} Paths in the complex $\tilde Z$- and 
  $\tilde \zeta$-planes for the curves shown in Fig.~\ref{deSitter}, where we
  hold the point $\tilde x$ fixed at O while moving $\tilde x'$ around the
  $(2+1)$ de Sitter space. If $Z<-1$, then $\tilde\zeta$ is purely imaginary.
  If $-1<Z<1$, $\tilde \zeta$ is essentially real (if it were not for the
  small $i\epsilon$ imaginary part). In this case, if $\chi_R<\chi_R'$,
  $-\pi<\tilde \zeta<0$, but if $\chi_R>\chi_R'$ then $0<\tilde\zeta<\pi$. The
  coincidence limit corresponds to both $\tilde\zeta=\pm \pi$, depending on
  that we approach $\tilde x$ from ``abov'' or ``below''. When $\tilde x$ and
  $\tilde x'$ are timelike related, $\tilde\zeta$ has both imaginary and real
  parts. Its real part is $\pm\pi$ depending on whether $\chi_R$ is greater or
  less than $\chi_R'$, respectively.}}
\end{quote}
\end{figure}

The ``supercurvature'' contribution of the $p^2=-1$ mode is in fact divergent.
This is related to the zero mode problem of massless quantum fields in
spacetimes with compact Cauchy surfaces. Following the usual prescription
\cite{zeromodeA,zeromodeB,zeromodeC,jaumekirsten}, we formally write this
divergent term as a divergent piece plus a finite one,
\begin{equation}
G^{(+)}_{\rm sup}(x,x') =  \frac{1}{4\pi^2}\langle0|Q^2|0\rangle + 
\sum_{l>0,m}\frac{1}{R_0^2}{\mathcal Y}_{-1,lm}\overline{{\mathcal
  Y}_{-1,lm}} , 
\end{equation}
where the infinity has been hidden in an infinite constant (see
\cite{jaumekirsten} for details). After, when taking derivatives to
compute the energy-momentum tensor, this divergent constant term will
give no contribution.  The sum can be performed, and the result is
\begin{align}
 G^{(+)}_{\rm sup}(x,x') &=\frac{1}{4\pi^2}\langle0|Q^2|0\rangle +
 \frac{1}{8\pi^2R_0^2}\left(
- 2\tilde\zeta\cot\tilde\zeta\right.\nonumber\\&+\left.
 (2\chi+i\pi)\tanh\chi_R+(2\chi'-i\pi)\tanh\chi'_R\frac{}{}\right),
\end{align}
where we have dropped an irrelevant constant.

Adding the continuum and supercurvature contributions, and
symmetrizing the result with respect to $x$ and $x'$, we finally find
the symmetric Hadamard two-point function (for pairs of points $x$,
$x'$ outside the lightcone in the ``left'' Minkowski patch),
\begin{align}
G^{(1)}(x,x') &=\frac{1}{4\pi^2\sigma}+\frac{1}{4\pi^2R_0^2}
\left(2\chi_R\tanh\chi_R+
2\chi'_R\tanh\chi'_R\nonumber\right.\\ &- \left.2\,\cot\tilde\zeta\,\mbox{arc
  tan}\frac{\sin\tilde\zeta}{\cos\tilde\zeta+\xi_M\xi_M'}\nonumber\right.\\
&+\left. \log\frac{(\xi_M\xi'_M)^2}
{\sin^2\tilde\zeta+(\cos\tilde\zeta+\xi_M\xi'_M)^2}\right)+
\frac{1}{2\pi^2}\langle0|Q^2|0\rangle \nonumber\\
&= \frac{1}{4\pi^2\sigma}+W(x,x'),
\end{align} 
where 
\begin{equation}
\sigma=\frac{1}{2}\left(\xi_R^2+\xi'_{R}{}^2+
2\xi_R\xi'_R\cos\tilde\zeta\right)
\end{equation}
is one half of the square geodesic distance in flat spacetime. The first term
in the final expression for $G^{(1)}(x,x')$ is the usual Minkowski ultraviolet
divergence.  The second term, $W(x,x')$, is due to the nontrivial geometric
boundary conditions imposed by the symmetry of our problem. If $W(x,x')$ were
not singular, the state would be of the Hadamard
form\cite{hadamardA,hadamardB,hadamardC}. But $W(x,x')$ has local and
nonlocal singularities. In the coincidence limit, it is divergent on the
bubble wall. It is logarithmically singular whenever one of the points is on
the lightcone emanating from the origin. It is also singular when $x$ and $x'$
satisfy the relation $\sin^2\tilde\zeta+(\cos\tilde\zeta+\xi_M\xi'_M)^2=0$, so
the argument of the logarithm diverges. The roots of this equation are at
\begin{equation}
\xi_M^{\rm s}{\xi'}^{\rm s}_M=-e^{\pm i\tilde\zeta_{\rm s}}.\label{rootsing}
\end{equation}
To clarify the position of the singularities, let us fix the point
$x_{\rm s}$ and look for the points $x'_{\rm s}$ which make
$G^{(+)}(x_{\rm s},x'_{\rm s})$ singular. Taking into account that
$\xi_M^{\rm s}$ and ${\xi'}^{\rm s}_M$ should be real and should
satisfy $0<\xi_M^{\rm s},{\xi'}^{\rm s}_M<1$, it is seen from
(\ref{rootsing}) that the allowed values of $\tilde\zeta_{\rm s}$ are
of the form $\pm\pi+iy$ (i.e., $\tilde x_{\rm s}$ and $\tilde x'_{\rm
  s}$ are ``timelike'' separated on a (2+1) de Sitter hyperboloid, see
Fig.~\ref{Zplane}), with $y>-\log\xi_M^{\rm s}>0$.  Without loss of
generality, we can assume that $\tilde x_{\rm s}= (0,0,0)$.  Then
$\cos\tilde\zeta_{\rm s}=-\cosh y=-\cosh {\chi_R'}^{\rm
  s}\cos\theta'_{\rm s}$. This implies that $0\leq\theta'_{\rm
  s}\leq\pi/2$, and we have no restriction on $\varphi'_{\rm s}$.
Since $\cosh {\chi_R'}^{\rm s} = \cosh y/\cos\theta'_{\rm s}\geq \cosh
y$, we find that ${\chi_R'}^{\rm s}\geq y\geq-\log\xi_M^{\rm s}$ or
${\chi_R'}^{\rm s}\leq-y\leq\log\xi_M^{\rm s}$.  So the region inside
of which (for any value of $\Omega'$) $G^{(+)}(x,x')$ is non singular
(apart from the singular points on the lightcone from $x$) is limited
by the curves
\begin{figure}[t]
\centering
\leavevmode\epsfysize=6cm \epsfbox{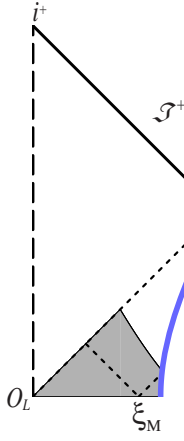}\\[3mm]
\begin{quote}
{\sl 
\caption[fig6]{\label{singularities} Nonlocal singularities in the upper
  half ``left Minkowski'', in the VIS model. Inside the shadowed region the
  two-point function $G^{(1)}(x,x')$, considered as a function of $x'$ with
  $x$ fixed, is singular only on the lightcone from $x$.}}
\end{quote}
\end{figure}
\begin{align}
{\xi'}^{\rm n.s.}_M&=\frac{1}{e^y\xi_M^{\rm s.}}\label{singularA}\\
{\chi'}^{\rm n.s.}_R &= \pm y\label{singularB},
\end{align} 
with $y>-\log\xi_M^{\rm s}$, the lightcone from the origin and the
bubble wall (the superscript n.s. stands for ``nearest (nonlocal)
singularity'', see Fig.~\ref{singularities}).  Note that as $x$
approaches to the bubble wall (i.e., $\xi_M\to 1$), the distance to
the nearest singular point $x'$ reduces.  Consistently, in the
limiting case when $x$ is on the wall, $W(x,x')$ is singular on the
coincidence limit.

Let us now check the causal relationship between singular points
satisfying equation (\ref{rootsing}). If we compute $\sigma(x_{\rm s},x'_{\rm
  s})$, we will find
\begin{equation}
\sigma(x_{\rm s},x'_{\rm s})=\frac{e^{-2
  y}}{2\xi_M^{{\rm s}\,2}}(1-\xi_M^{{\rm s}\,2})
(1-\xi_M^{{\rm s}\,2} e^{2y})\leq 0,
\end{equation}
where the last inequality follows from $y\geq-\log\xi_M^{\rm s}$,
$0\leq\xi_M^{\rm s}\leq 1$.  The equality can only be realized if $x_{\rm
  s}$ is on the bubble wall.  Then, in this case, there exist nonlocal
singularities (of $W(x,x')$) which are null related. But if $x_{\rm s}$ is
not on the bubble wall, its singular partners are always time-like related
with it. 

Summarizing, the two-point function is locally Hadamard everywhere except on
the bubble wall and on the lightcone. Moreover, it has (harmless, see
discussion below) nonlocal singularities.

\subsection{ST model}
In this second model, which has previously been considered by Sasaki
and Tanaka \cite{TMdecay}, the $\widehat\Phi$ field interacts with the
tunneling field $\sigma$ only on the bubble wall. We assume the
infinitely thin-wall approximation, so the interaction term can be
written as
\begin{equation}
m^2(\eta) = 2 \frac{V_0}{R_0^2} \delta(\eta),
\end{equation}
where $V_0>0$ characterizes the strength of the interaction and $R_0$ is the
radius of the bubble wall.

Decomposing the field $\widehat{\bf \Phi}$ as before, we find that the
Schr{\"o}dinger equation for $F_p$ takes the form
\begin{equation}
-F''_p+2 V_0\delta(\eta)F_p = p^2 F_p,
\end{equation}
Now the spectrum is purely
continuous with $p^2>0$. The solution of this Schr{\"o}dinger equation is the
scattering basis (\ref{scatt}) with the transmission and reflection
coefficients given by
\begin{align}
\rho(p) &= \frac{V_0}{i|p|-V_0},\\
\sigma(p) &= \frac{i|p|}{i|p|-V_0}.  
\end{align}
Following a similar path\footnote{Detailed computations
  will be presented elsewhere\cite{next}}, we arrive at the following
Hadamard two-point function (for points $x$, $x'$ in the Rindler wedge
and inside the bubble),
\begin{align}
G^{(1)}(x,x')&=\frac{1}{4\pi^2\sigma}+
\frac{1}{4\pi^2 i}\frac{1}{\xi_R\xi'_R\sin\tilde\zeta}
\left(_2F_1\left[1,V_0;V_0+1;-e^{i\tilde\zeta}
\frac{\xi_R\xi'_R}{R_0^2}\right]\right.\nonumber\\
&- \left._2F_1\left[1,V_0;V_0+1;-e^{-i\tilde\zeta}
\frac{\xi_R\xi'_R}{R_0^2}\right]\right)=\frac{1}{4\pi^2\sigma}+W(x,x'),
\end{align}
which is explicitly SO(3,1) invariant. If we take the coincidence limit, the
function $W(x,x')$ has divergences on the bubble wall, so the state is not
locally Hadamard. Apart from this, it has nonlocal logarithmic singularities
at the points where the argument of the hypergeometric functions become 1,
i.e., whenever $\xi_M\xi_M'=-\exp(\pm i\tilde \zeta)$. This is the same
relation we found in the VIS model. Borrowing the conclusions from the VIS
model, the state is locally Hadamard everywhere except on the bubble wall, and
has (harmless) timelike nonlocal singularities (except also on the
bubble wall).

As we have seen, the two models we have considered share two singular
behaviors: the existence of nonlocal singularities and the singularity of
$W(x,x')$ in the coincidence limit on the bubble wall. These singularities
seem to be related with the oversimplification of the model.  Presumably, if
instead of a $\delta$-like term interaction we had introduced a smooth
function, these divergences would disappear.

\section{Renormalized expectation value of the stress tensor}

As we have pointed out, for the two models we have studied the singularities
of $G^{(1)}(x,x')$ are nearly of the Hadamard type.  We can use the
point-splitting regularization prescription to compute the renormalized
expectation value of the stress-energy momentum tensor
\cite{christensen,waldreg,folacci},
\begin{align}
\langle T_{ab}\rangle &= \frac{1}{2}\lim_{x\to x'}{\mathcal
  D}_{ab'}[W(x,x')],\\
{\mathcal D}_{ab'} &= \nabla_a\nabla_{b'}-
\frac{1}{2}g_{ab'}g^{cd'}\nabla_c\nabla_{d'}.
\end{align} 
Noticing that $\cos\tilde\zeta =
-Z+i\epsilon=-\cos\sqrt{2\,\,^{dS}\sigma}+i\epsilon$, where $^{dS}\sigma$ is
one half of the square distance in a unit (2+1) de Sitter spacetime, the
covariant derivatives in the ``de Sitter'' direction are easily computed from
\cite{christensen,hadamardC}
\begin{align}
[^{dS}\sigma_{;A}] &=0, \\
{[^{dS}\sigma_{;AB'}]} &= -\frac{g_{AB}}{\xi_R^2},
\end{align} 
where the brackets stand for the coincidence limit.

\subsection{VIS model}

The renormalized expectation value of the stress tensor turns out to be
\begin{align}
\langle T_{\xi_R\xi_R}\rangle&=\frac{\xi_R^2-2 R_0^2}{4\pi^2 R_0^2 
(R_0^2-\xi_R^2)^2}, \\
\langle T_{AB}\rangle &= -\frac{\xi_R^4-3 R_0^2\xi_R^2+6 R_0^4}
{12\pi^2R_0^2(R_0^2-\xi_R^2)^3}\,g_{AB}.
\end{align}
where $0\leq\xi_R\leq R_0$ (i.e., $x$ is in the left Rindler wedge and
inside the bubble).
It is clear from the
expression that the energy-momentum tensor behaves somewhat better than the
two-point function. It is divergent on the bubble wall, but behaves smoothly
on the lightcone. So it can be analytically continued to the inside of the
lightcone. There it behaves like the energy-momentum tensor
of a perfect fluid,
\begin{align}
\langle T_{ab}\rangle = (\rho + p)u_a u_b + p g_{ab},
\end{align}
where $u^a=(\partial_t)^a$. On the lightcone it satisfies the equation
of state $p=-\rho$ whith
\begin{align}
\rho=\frac{1}{2\pi R_0^4}
\end{align}
For large $t$ the equation of state turns out to be
\begin{equation}
p=-\frac{1}{3}\rho,
\end{equation}
with
\begin{equation}
\rho=\frac{1}{4\pi^2}\frac{1}{R_0^2t^2}.
\end{equation}
Taking into account that the field is massless except on the
bubble-wall, one might naively expect that the energy momentum tensor
would behave like radiation, with $\rho\propto t^{-4}$. Instead of
this, we have found that it decreases slower. In fact, it can be shown
that its behaviour is dominated by gradients of the supercurvature
modes\footnote{The particle content and interpretation of the vacuum we
  have considered will be discussed
  elsewhere\cite{next}\label{particle}}.

\subsection{ST model}
For the ST model, we find
\begin{align}
\langle T_{\xi_R\xi_R}\rangle &= \frac{1}{2\pi^2R_0^4}
\frac{V_0}{V_0+2}{}_2F_1\left[3,V_0+2;V_0+3;
\left(\frac{\xi_R}{R_0}\right)^2\right],\\
 \langle T_{AB}\rangle &= \frac{1}{2\pi^2R_0^4}\frac{V_0}{V_0+2}\left(
_2F_1\left[3,V_0+2;V_0+3;
\left(\frac{\xi_R}{R_0}\right)^2\right]\nonumber\right.\\
&+\left.2\left(\frac{\xi_R}{R_0}\right)^2
\frac{V_0+2}{V_0+3}{}_2F_1\left[4,V_0+3;V_0+4;
\left(\frac{\xi_R}{R_0}\right)^2\right]\right)g_{AB},
\end{align}
where $0\leq\xi_R\leq R_0$.  As before, the energy-momentum tensor
turns out to be singular only on the bubble wall\footnote{The quantum
  state found in\cite{TMdecay} has the problem of being ill defined on
  the light-cone. This singularity propagates to the renormalized
  energy-momentum tensor, making it to blow up on the light cone.
  This seems to be due to an inappropiate normalization of the mode
  functions.}.  Continuing analytically the results to the inside of
the light-cone, we find that it is of the perfect fluid form.
On the lightcone it satisfies the equation of state $p=-\rho$ with
\begin{align}
\rho=-\frac{1}{2 \pi^2 R_0^4}\frac{V_0}{V_0+2}.
\end{align}
For large $t$ the equation of state turns
out to be
\begin{equation}
p=\rho,
\end{equation}
with
\begin{equation}
\rho = -\frac{4\,R_0^2}{t^6}.
\end{equation}
 Notice that in this model the energy density is negative and
decreases faster than radiation\footnotemark[5].

\section{Discussion}
In this paper we have performed the computation of $\langle T_{ab}\rangle$ in
a quantum state which fulfills our basic requirement of SO(3,1) invariance.
In fact, we have just outlined the most simple method to find a SO(3,1)
invariant state.  The question is whether by choosing a different set of modes
we can also obtain an inequivalent SO(3,1) invariant state but also of the
Hadamard form. A theorem due to Kay and Wald \cite{waldkay} is illuminating in
this respect.  The theorem states that in a spacetime with a bifurcate Killing
horizon there can exist at most one regular quasifree state invariant under
the isometry which generates the bifurcate Killing horizon.  Let us briefly
analyze the conditions under which the theorem holds.

In (3+1) spacetimes, we get a bifurcate Killing horizon whenever a one
parameter group of isometries leaves invariant a 2-dimensional spacelike
manifold $\Sigma$.  The bifurcate Killing horizon is generated by the null
geodesics orthogonal to $\Sigma$\cite{waldkay}. For example, Minkowski
spacetime has bifurcate Killing horizons. The isometry group is a
one-parameter subgroup of Lorentz boosts, and the manifold $\Sigma$ is a
two-plane. Any SO(3,1) invariant spacetime, where the line element can be
written in the form
\begin{equation}
ds^2 = d\xi_R^2 + a(\xi_R)^2(-d\chi_R^2+\cosh^2\chi_Rd\Omega),
\end{equation} 
has a SO(3,1) invariant bifurcate Killing horizon. Noticing that the
$\xi_R=\mbox{ct.}$ hypersurfaces are (2+1) de Sitter spaces which can be
thought as embedded in a (3+1) Minkowski space, any boost generator on these
hypersurfaces is the infinitesimal generator of a isometry which 1) leaves
invariant a spacelike 2-manifold (so we get a bifurcate Killing horizon) and
2) leaves any SO(3,1) symmetric state invariant. We can take, for example,
the boost generator in the $ZT$ plane of the embedding Minkowski space.
Expressed in the Rindler coordinates, it becomes
\begin{equation}
\xi^a = -\cos\theta\frac{\partial}{\partial\chi_R} + \tanh\chi_R\sin\theta
\frac{\partial}{\partial\theta}.
\end{equation}
The Killing field $\xi^a$ leaves invariant the spacelike 2-manifold
$\theta=\pi/2$, $\chi_R=0$. All bubble spacetimes with or without the
inclusion of gravity do possess this bifurcate Killing horizon.

A (pure) quasifree ground state is the (wider) algebraic version (see
\cite{waldkay} and references therein) of what is usually called a
``frequency splitting'' Fock vacuum state. A quasifree state has the
special property of being completely characterized from its two-point
function. A regular\footnote{We include the notion of globally
  Hadamard in the definition of a regular state.} quasifree ground
state is a quasifree ground state whose two-point symmetric function
is globally Hadamard and which has no zero modes. The VIS model has a
zero mode (as any massless field in spacetimes with compact Cauchy
surfaces have \cite{waldkay}), so the theorem cannot be directly
applied. Also strictly speaking, the quantum state we have found for
the ST model does not fulfill the requirements of the theorem, because
it is not globally Hadamard.  Roughly speaking, a two-point function
is said to be globally Hadamard if it is locally Hadamard and in
addition has nonlocal singularities only at points $x$, $x'$ which are
null related within a causal normal neighborhood of a Cauchy
hypersurface\footnote{A more rigorous definition of globally Hadamard
  states can be found in \cite{waldkay}}. As we have seen, if we
ignore the problems on the bubble wall, the Hadamard function
$G^{(1)}(x,x'$) we have found for the ST model has nonlocal
singularities, but they are timelike related.  So, if it were not for
the singularities on the bubble wall, the state would be globally
Hadamard and without zero modes.  As stated before, we think that the
singularities on the bubble wall would disappear if the potential were
modeled by a smooth function instead of by a $\delta$-like term,
making the state globally Hadamard.  Then, symmetry would suffice to
determine the (physically admissible) quantum state for this model.
Generic models which would not present these pathologies will be
presented elsewhere\cite{next}

\section*{Acknowledgements}

I would like to thank Edgard Gunzig for his kind hospitality at the
Peyresq-3 meeting. I am grateful to Jaume Garriga for many helpful
discussions.  I acknowledge support from European Project
CI1-CT94-0004, and from CICYT under contract AEN98-1093.


\begin{thebibliography}{10}

\bibitem{coleman}
{S. Coleman, Phys. Rev. D{\bf 15}, 2929 (1977); C.G. Callan and S. Coleman,
  {\sl ibid.} {\bf 16}, 1762 (1977)}.

\bibitem{rubakov}
{V.A. Rubakov, Nucl. Phys. {\bf B245}, 481 (1984).}

\bibitem{VV}
{T. Vachaspati and A. Vilenkin, Phys. Rev. D{\bf 43}, 3846 (1991)}.

\bibitem{MT}
{T. Tanaka and M. Sasaki, Phys. Rev. D{\bf 49}, 1039 (1994); T. Tanaka and M.
  Sasaki, {\sl ibid.} {\bf 50}, 6444 (1994).}

\bibitem{TMdecay}
{M. Sasaki, T. Tanaka, K. Yamamoto, and J. Yokoyama, Progr. Theor. Phys. {\bf
  90}, 1019 (1993)}.

\bibitem{colemanmin}
{S. Coleman, V. Glaser and A. Martin, Commun. Math. Phys. {\bf 58}, 211
  (1978)}.

\bibitem{banks}
{T. Banks, C.M. Bender and T. T. Wu, Phys. Rev. D{\bf 8}, 3346(1973); T. Banks
  and C. M. Bender, {\sl ibid.} {\bf 8}, 3366(1973)}.

\bibitem{vega}
{H. J. Vega, J. L. Gervais, and B. Sakita, Nucl. Phys. {\bf B139}, 20 (1978);
  {\sl ibid.} {\bf B142}, 125 (1978); Phys. Rev. D{\bf 19}, 604 (1979)}.

\bibitem{gencol}
S.~Coleman.
\newblock {\em Aspects of Symmetry}.
\newblock Cambridge University Press, Cambridge, 1985.

\bibitem{hadamardA}
F.G. Friedlander.
\newblock {\em The Wave Equation on a Curved Spacetime}.
\newblock Cambridge University Press, Cambridge, 1975.

\bibitem{hadamardB}
J.~Hadamard.
\newblock {\em Lectures on Cauchy's problem in Linear Partial Differential
  Equations}.
\newblock Yale University Press, New Heaven, CT, 1923.

\bibitem{hadamardC}
B.S. DeWitt.
\newblock {\em Relativity, Groups and Topology}.
\newblock Gordon and Breach, New York, 1964.

\bibitem{waldkay}
{B. S. Kay and R.M. Wald, Phys. Rep. {\bf 207}, 49 (1991); B. S. Kay J. Math.
  Phys. {\bf 34}, 4519 (1993)}.

\bibitem{christensen}
{S. M. Christensen, Phys. Rev. D{\bf 14}, 2490 (1974); S. M. Christensen, Phys.
  Rev. D{\bf 17}, 946 (1978) }.

\bibitem{waldreg}
{R.M. Wald, Commun. Math. Phys. {\bf 54}, 1 (1977); R.M. Wald, Phys. Rev. D{\bf
  17}, 1477 (1978)}.

\bibitem{folacci}
{M. R. Brown and A. Ottewill, Phys. Rev. D{\bf 34}, 1776 (1986); D. Bernard and
  A. Folacci, {\sl ibid.} {\bf 34}, 2286 (1986)}.

\bibitem{dorca}
{M. Dorca and E. Verdaguer, Nucl. Phys. {\bf B484} 435 (1997)}.

\bibitem{VIS}
{A. Vilenkin Phys. Lett. {\bf 133B}, 177 (1983); J. Ipser and P. Sikivie, Phys.
  Rev. D{\bf 30}, 712 (1984)}.

\bibitem{bunchdavies}
{T. S. Bunch and P. C. W. Davies, Proc. R. Soc. London {\bf A360}, 117 (1978)}.

\bibitem{zeromodeA}
B.S. DeWitt.
\newblock {\em Relativity, Groups and Topology II}.
\newblock North Holland, Amsterdam, 1983.

\bibitem{zeromodeB}
{L.H. Ford and C. Pathinayake, Phys. Rev. D{\bf 39}, 3642 (1989)}.

\bibitem{zeromodeC}
{J. Garriga and A. Vilenkin, Phys. Rev. D{\bf 45}, 3469 (1993)}.

\bibitem{jaumekirsten}
{K. Kirsten and J. Garriga, Phys. Rev. D{\bf 48}, 567 (1993)}.

\bibitem{sumaY}
{P. Candelas and D. J. Raine, Phys. Rev. D{\bf 12}, 965 (1975); B. Ratra, Phys.
  Rev. D{\bf 31}, 1931 (1985) }.

\bibitem{next} X. Montes, in preparation.

\end{thebibliography}
\end{document}